%% file: paper.tex
\documentclass[11pt, a4paper]{article}

\input{my-macros}

\title{\LARGE \bf Kernel Methods for Nonlinear Connectivity Detection}
\author{Lucas Massaroppe 
\thanks{Institudo de Astronomia, Geof\'{i}sica e Ci\^{e}ncias Atmosf\'{e}ricas,
		University of S\~{a}o Paulo, S\~{a}o Paulo, Brazil. 
		Corresponding author: {\tt\small lucasmassaroppe@usp.br}.
		This work was supported in part by an institutional Ph.D. CAPES Grant at Escola Polit\'{e}cnica, University of S\~{a}o Paulo, Brazil.}%
\and
		Luiz A. Baccal\'{a} %
\thanks{Escola Polit\'{e}cnica,
        University of S\~{a}o Paulo, Brazil.}%
}%

\begin{document}
\maketitle


\begin{abstract}
	
	In this paper, we show that the presence of nonlinear coupling between time series may be detected employing kernel feature space representations alone dispensing with the need to go back to solve the \textit{pre-image problem} to gauge model adequacy. As a consequence, the canonical methodology for model construction, diagnostics, and Granger connectivity inference applies with no change other than computation using kernels \textit{in lieu} of second-order moments.
	
	
\end{abstract}


\section{Introduction}
\label{sec:introduction}

\input{introduction}


\section{Problem Formulation}
\label{sec:prob_form}

\input{prob_form}


\section{Numerical Illustrations}
\label{sec:num_ill}

\input{num_ill}


\section{Conclusion and Future Work}
\label{sec:disc}

\input{disc}


\section*{Acknowledgement}
\label{sec:ack}

L.M. gratefully acknowledges support from an institutional Ph.D. CAPES Grant. L.A.B. also gratefully acknowledges to CNPq Grant $308073/2017-7$.


\section*{Remark}
\label{sec:discl}

This preliminary version is being prepared for submission to a refereed journal.


\bibliographystyle{abbrv}
\bibliography{references}

\end{document}

%% file: my-macros.tex
\usepackage[latin1]{inputenc}
\usepackage[english]{babel}
\usepackage{amsmath, amscd, amsthm, amssymb, amsfonts}
\usepackage{makeidx}
\usepackage[left=2cm,right=2cm,top=2cm,bottom=2cm]{geometry}
\usepackage{pstool}
\usepackage{subfig}
\usepackage{epsfig}
\usepackage{graphicx,psfrag}
\usepackage{trfsigns}
\usepackage{cite}
\usepackage{tikz}
\usetikzlibrary{shapes,shadows,calc}
\usepgflibrary{arrows}

\usepackage{url}
\usepackage[colorlinks=true, citecolor=blue, breaklinks=true]{hyperref}

\hypersetup{
    bookmarks=true,         
	bookmarksopen=true,
	bookmarksnumbered=true,
    unicode=true,           
    pdftoolbar=true,        
    pdfmenubar=true,        
    pdffitwindow=true,      
    pdfstartview={FitH},    
    pdftitle={Kernel Methods for Nonlinear Connectivity Detection},  
    pdfauthor={L.M. and L.A.B.},       
    pdfsubject={arXiv paper},          
    pdfcreator={Lucas Massaroppe},     
    pdfproducer={TextMate (macOS)},    
    pdfkeywords={Nonlinear time series} {Nonlinear-Granger causality} {Inference}, 
    pdfnewwindow=true,       
	colorlinks=true,         
	breaklinks=true
}

\DeclareMathOperator{\rank}{rank}	 
\DeclareMathOperator{\vect}{vec}     
\DeclareMathOperator{\E}{\mathbb{E}} 

\let\oldsqrt\sqrt
\def\sqrt{\mathpalette\DHLhksqrt}
\def\DHLhksqrt#1#2{%
\setbox0=\hbox{$#1\oldsqrt{#2\,}$}\dimen0=\ht0
\advance\dimen0-0.2\ht0
\setbox2=\hbox{\vrule height\ht0 depth -\dimen0}%
{\box0\lower0.4pt\box2}}

\makeatletter
\let\save@mathaccent\mathaccent
\newcommand*\if@single[3]{%
  \setbox0\hbox{${\mathaccent"0362{#1}}^H$}%
  \setbox2\hbox{${\mathaccent"0362{\kern0pt#1}}^H$}%
  \ifdim\ht0=\ht2 #3\else #2\fi
  }
\newcommand*\rel@kern[1]{\kern#1\dimexpr\macc@kerna}
\newcommand*\widebar[1]{\@ifnextchar^{{\wide@bar{#1}{0}}}{\wide@bar{#1}{1}}}
\newcommand*\wide@bar[2]{\if@single{#1}{\wide@bar@{#1}{#2}{1}}{\wide@bar@{#1}{#2}{2}}}
\newcommand*\wide@bar@[3]{%
  \begingroup
  \def\mathaccent##1##2{%
    \let\mathaccent\save@mathaccent
    \if#32 \let\macc@nucleus\first@char \fi
    \setbox\z@\hbox{$\macc@style{\macc@nucleus}_{}$}%
    \setbox\tw@\hbox{$\macc@style{\macc@nucleus}{}_{}$}%
    \dimen@\wd\tw@
    \advance\dimen@-\wd\z@
    \divide\dimen@ 3
    \@tempdima\wd\tw@
    \advance\@tempdima-\scriptspace
    \divide\@tempdima 10
    \advance\dimen@-\@tempdima
    \ifdim\dimen@>\z@ \dimen@0pt\fi
    \rel@kern{0.6}\kern-\dimen@
    \if#31
      \overline{\rel@kern{-0.6}\kern\dimen@\macc@nucleus\rel@kern{0.4}\kern\dimen@}%
      \advance\dimen@0.4\dimexpr\macc@kerna
      \let\final@kern#2%
      \ifdim\dimen@<\z@ \let\final@kern1\fi
      \if\final@kern1 \kern-\dimen@\fi
    \else
      \overline{\rel@kern{-0.6}\kern\dimen@#1}%
    \fi
  }%
  \macc@depth\@ne
  \let\math@bgroup\@empty \let\math@egroup\macc@set@skewchar
  \mathsurround\z@ \frozen@everymath{\mathgroup\macc@group\relax}%
  \macc@set@skewchar\relax
  \let\mathaccentV\macc@nested@a
  \if#31
    \macc@nested@a\relax111{#1}%
  \else
    \def\gobble@till@marker##1\endmarker{}%
    \futurelet\first@char\gobble@till@marker#1\endmarker
    \ifcat\noexpand\first@char A\else
      \def\first@char{}%
    \fi
    \macc@nested@a\relax111{\first@char}%
  \fi
  \endgroup
}
\makeatother

\makeatletter
\def\CatchFBT@Fin@l#1[#2]{%
   \begingroup
      \makeatletter #2%
      \scantokens\expandafter{%
         \expandafter\CatchFBT@tok\expandafter{\the\CatchFBT@tok}}%
      \CatchFBT@IsAToken{#1}
         {\global#1\expandafter{\the\CatchFBT@tok}}
         {\xdef#1{\the\CatchFBT@tok}}%
      \ifx\CatchFBT@tok#1\else\global\CatchFBT@tok{}\fi
   \endgroup
}
\makeatother

%% file: introduction.tex
Describing `connectivity' has become of paramount interest in many areas of investigation that involve interacting systems. Physiology, climatology, and economics are three good examples where dynamical evolution modelling is often hindered as system manipulation may be difficult or unethical. Consequently, interaction inference is frequently constrained to using time observations alone.

A number of investigation approaches have been put forward \cite{Schumacker2004, Applebaum2008, Cover2006, Hlavackovaschindler2007, Schreiber2000}. However, the most popular and traditional one still is the nonparametric computation of cross-correlation (CC) between pairs of time series, and variants thereof, like coherence analysis \cite{Bendat1980}, even despite their many shortcomings \cite{Baccala2001a}.

When it comes to connectivity analysis, recent times have seen the rise of Granger Causality (GC) as a unifying concept. This is mostly due to GC's unreciprocal character \cite{Granger1969} (as opposed to CC) which allows establishing the direction of information flow between component subsystems.

Most GC approaches rest on fitting parametric models to time series data and, again as opposed to CC, under appropriate conceptualization, also holds for more than just pairs of time series, giving rise to the ideas of (a) Granger connectivity and (b) Granger influentiability \cite{Baccala2014epi}. 

GC inferential methodology is dominated by the use of \textit{linear} multivariate time series models \cite{Lutkepohl2005}. This is so because linear models have statistical properties (and shortcomings) that are well understood besides having the advantage of sufficing when the data are Gaussian. As an added advantage GC characterization allows immediate frequency domain connectivity characterization via concepts like `directed coherence' (DC) and `partial directed coherence' (PDC) \cite{Baccala2001}. 

It is often the case, however, that data gaussianity does not hold. Whereas nonparametric approaches do exist \cite{Schumacker2004, Hlavackovaschindler2007, Schreiber2000}, parametric nonlinear modelling offers little relief from the need for long observation data sets for reliable estimation in sharp contrast to linear models that perform well under the typical scenario of fairly short datasets over which natural phenomena can be considered stable. A case in point is neural data where animal behaviour changes are associated with relatively short-lived episodic signal modifications. 

The motivation for the present development is that reproducing kernel transformations applied to data, as in the support vector machine learning classification case \cite{Vapnik1998}, can effectively produce estimates that inherit many of the good convergence properties of linear methods. Because they carry over these properties under proper kernelization, it is possible to show that nonlinear links between subsystems can be rigorously detected.

In Section \ref{sec:prob_form} we formulate the problem and review some background about reproducing kernel theory together with the main results which are backed up by extensive numerical Monte Carlo illustrations in Section \ref{sec:num_ill}. Conclusions and current problem status and challenges end the paper (Section \ref{sec:disc}).

%% file: prob_form.tex
The most popular approach to investigating GC connectivity is through modeling multivariate time series via linear vector autoregressive models \cite{Lutkepohl2005}, where the central idea is to compare prediction effectiveness for a time series $x_i(n)$ when the past of other time series is taken into account in addition to its own past. Namely,
\begin{equation}
	\label{eq:lin_Var}
	\mathbf{x}(n)=\sum_{k=1}^p\mathbf{A}_k\mathbf{x}(n-k)+\mathbf{w}(n).
\end{equation}
Under mild conditions, \eqref{eq:lin_Var}  constitutes a valid representation of a linear stationary stochastic process where the evolution of $\mathbf{x}(n)=[x_1(n), \cdots, x_D(n)]^{\top}$ is obtained by filtering suitable $\mathbf{w}(n)=[w_1(n), \cdots, w_D(n)]^{\top}$ \textit{purely stochastic innovation} processes, i.e. where $w_i(n)$ and $w_j(m)$ are independent provided $n \neq m$ \cite{Priestley1981}. If $w_i(n)$ are jointly Gaussian, so are $x_i(n)$ and the problem of characterizing connectivity reduces to well known procedures to estimate the $\mathbf{A}_k$ parameters in \eqref{eq:lin_Var} via least squares, which is the applicable maximum likelihood procedure. Nongaussian $w_i(n)$ translate into nongaussian $x_i(n)$ even if some actual \eqref{eq:lin_Var} linear generation mechanism holds. Linearity among  nongaussian $x_i(n)$ time series may be tested with help of cross-polyspectra \cite{Nikias1993, SubbaRao1984}, which, if unrejected, still allows for a representation like \eqref{eq:lin_Var} whose optimal estimation requires a suitable likelihood function to accommodate the observed nongaussianity.

If linearity is rejected, $x_i(n)$ nongaussianity is a sign of nonlinear mechanisms of generation, even if
\begin{equation}
	\label{eq:gennonlin}
	\mathbf{x}(n)=\mathbf{g}(\mathbf{x}(n_{-}),\mathbf{w}(n)),
\end{equation}
which generalizes \eqref{eq:lin_Var} where $\mathbf{x}(n_{-})$ stands for $\mathbf{x}(n)$'s past under some suitable dynamical law $\mathbf{g}(\cdot)$.

The distinction between (a) nonlinear $x_i(n)$ that are nonetheless linearly coupled as in \eqref{eq:lin_Var} under nongaussian $\mathbf{w}(n)$ and (b) fully nonlinearly coupled  processes is often overlooked. In the former case, linear methods suffice for connectivity detection \cite{Schelter2006} but fail in the latter case \cite{Massaroppe2011a} calling for the adoption of alternative approaches. In some cases, however, linear approximations are inadequate in so far as to preclude connectivity detection \cite{Massaroppe2011a}.

In the present context, solution to the connectivity problem entails a suitable data driven approximation of $ \mathbf{g}(\cdot)$ whilst singling out the $x_i(n)$ and $x_j(n)$ of interest. To do so we examine the employment of \textit{kernel} methods \cite{Scholkopf2002} where functional characterization is carried out by with help of a high dimensional space representation
\begin{equation}
	\label{eq:Kernel_repre}
	\boldsymbol{\phi}: \mathbb{X} \to \mathbb{F},
\end{equation}
for $F= \dim (\mathbb{F}) \gg D= \dim (\mathbb{X}$) where $\boldsymbol{\phi}(\mathbf{x}(n))$ is a mapping from the input space $\mathbb{X}$ into the feature space $\mathbb{F}$ whose role is to properly unwrap the data and yet insure that the inner product $\langle \boldsymbol{\phi}(\mathbf{x})|\boldsymbol{\phi}(\mathbf{y})\rangle$ can be written as a simple function of $\mathbf{x}$ and $\mathbf{y}$ dispensing with the need for computations in $\mathbb{F}$. This possibility is granted by chosing $\boldsymbol{\phi}(\mathbf{x})$ to satisfy the so called \textit{Mercer condition} \cite{Mercer1909}. 
 
A simple example of \eqref{eq:Kernel_repre} is the mapping
\begin{equation}
	\label{eq:kernel_ex}
	\phi : x \mapsto \langle \phi(x) |\;=\; [c ,\; \sqrt{2c x}, x^2]^{\top},
\end{equation} 
for $x\in \mathbb{R}$ and $\langle \phi(x) | \in \mathbb{F} $ using Dirac's bra-ket notation. In this case the Mercer kernel is given by
\begin{equation}
	\label{eq:polyn_kernel}
	\kappa(x,y)=\langle \boldsymbol{\phi}(x)|  \boldsymbol{\phi}(y)\rangle = (c+xy)^2,
\end{equation}
which is the simplest example of a polynomial kernel \cite{Scholkopf2002}.

In the multivariate time series case, we consider
\begin{equation}
	\boldsymbol{\phi}: \mathbf{x}(n) \mapsto [\langle \phi_1(x_1(n))|, \cdots, \langle \phi_i(x_i(n))|, \cdots, \langle \phi_D(x_D(n))|]^{\top},
\end{equation}
where, for simplicity, we adopt the same transformation $\phi (\cdot)=\phi_i(\cdot)=\phi_j(\cdot)$ for each $x_i(n)\in\mathbb{R}$ time series component so that the
\begin{equation}
	\label{eq:Scalar_in_F}
	\langle  \boldsymbol{\phi}(\mathbf{x}(n))|  \boldsymbol{\phi}(\mathbf{x}(m))\rangle=\mathbf{K}(\mathbf{x}(n),\mathbf{x}(m)),
\end{equation}
is a matrix whose elements are given by $K_{ij}(n,m)=\langle \phi(x_i(n))|\phi(x_j(m))\rangle$.

Rather than go straight into the statement of the general theory, a simple example is more enlightening. In this sense consider a bivariate stationary time series
\begin{eqnarray}
	x_1(n) & = & g_1(x_1(n-1),w_1(n)),         \label{eq:itself} \\
	x_2(n) & = &g_2(x_1(n-1),x_2(n-1),w_2(n)), \label{eq:another}
\end{eqnarray}
where $g_i(\cdot)$ are nonlinear functions and only the previous instant is relevant in producing the present behaviour. An additional feature, thru \eqref{eq:another}, is that $x_1(n)$ is connected to (Granger causes) $x_2(n)$ but not conversely. Application of the kernel transformation leads to,
\begin{eqnarray}
	\langle \phi(x_1(n))|&=&\langle \phi(g_1(x_1(n-1),w_1(n)))|,          \label{eq:itself_inF}  \\
	\langle \phi(x_2(n))|&=&\langle\phi( g_2(x_1(n-1),x_2(n-1),w_2(n)))|. \label{eq:another_inF}
\end{eqnarray}

However if one assumes the possibility of a linear approximation in $\mathbb{F}$ one may write
\begin{equation}
	\label{eq:kernel_lin_biv}
	\left[
	\begin{array}{c}
		\langle \phi(x_1(n))|\\
		\langle \phi(x_2(n))|
	\end{array}
	\right]
	=
	\left[
	\begin{array}{cc}
		\alpha_{11} & \alpha_{12}\\
		\alpha_{21} & \alpha_{22}
	\end{array}
	\right]
	\left[
	\begin{array}{c}
		\langle \phi(x_1(n-1))|\\
		\langle \phi(x_2(n-1))|
	\end{array}
	\right]
	+
	\left[
	\begin{array}{c}
		\langle \widetilde{w}_1(n)|\\
		\langle \widetilde{w}_2(n)|
	\end{array}
	\right],
\end{equation}
where $[\langle \widetilde{w}_1(n)| \; \langle \widetilde{w}_2(n)|]^{\top}$ stands for approximation errors in the form of innovations. Mercer kernel theory allows taking the external product with respect to $[|\phi(x_1(n-1)) \rangle \; |\phi(x_2(n-1)) \rangle]^\top$ on both sides of \eqref{eq:kernel_lin_biv} leading to
\begin{equation}
\label{eq:kernel_matrix_biv}
	\mathbf{K}(\mathbf{x}(n),\mathbf{x}(n-1)) = \mathbf{A}\;\mathbf{K}(\mathbf{x}(n-1),\mathbf{x}(n-1)),
\end{equation}
after taking expectations on both sides where
\begin{equation}
\label{eq:small_A}
	\mathbf{A} = 
	\left[\begin{array}{cc}
		\alpha_{11} & \alpha_{12}\\
		\alpha_{21} & \alpha_{22}
	\end{array}\right]
\end{equation}
and
\begin{equation}
	\label{eq:K_biv}
	\mathbf{K}(\mathbf{x}(n),\mathbf{x}(m)) =
	\left[
	\begin{array}{cc}
		\E[\langle \phi(x_1(n))|\phi(x_1(m))\rangle] & \E[\langle \phi(x_1(n))|\phi(x_2(m))\rangle] \\
		\E[\langle \phi(x_2(n))|\phi(x_1(m))\rangle] & \E[\langle \phi(x_2(n))|\phi(x_2(m))\rangle]
	\end{array}
	\right],
\end{equation}
since $\E[\langle \widetilde{w}_i(n)|\phi(x_j(m))\rangle]=0$ for $n>m$ given that $\langle \widetilde{w}_i(n)|$ plays a zero mean innovations role.

It is easy to obtain $\mathbf{A}$ from sample kernel estimates. Furthermore  it is clear that \eqref{eq:itself} holds if and only if $\alpha_{12}=0$.

To \eqref{eq:kernel_matrix_biv}, which plays the role of Yule-Walker equations, one may add the following equation to compute the innovations covariance
\begin{equation}
	\label{eq:sigma_w}
	\boldsymbol{\Sigma}_{\langle \widetilde{\mathbf{w}}(n)|}=\mathbf{K}(\mathbf{x}(n),\mathbf{x}(n))-\mathbf{A}\mathbf{K}(\mathbf{x}(n),\mathbf{x}(n))\mathbf{A}^{\top},
\end{equation}
\begin{equation}
	\label{eq:kernel_matrix_biv_simple}
	\mathbf{K}(-1) = \mathbf{A}\mathbf{K}(0),
\end{equation}
where only reference to the $m-n$ difference is explicitly denoted assuming signal stationarity. Also \eqref{eq:sigma_w} simplifies to
\begin{equation}
	\label{eq:sigma_w_simple}
	\boldsymbol{\Sigma}_{\langle \widetilde{\mathbf{w}}(n)|}=\mathbf{K}(0)-\mathbf{A}\mathbf{K}(0)\mathbf{A}^{\top}=\mathbf{K}(0)-\mathbf{K}(-1)\mathbf{A}^{\top}.
\end{equation}

This formulation is easy to generalize to model orders $p>1$ and to more time series via
\begin{equation}
	\label{eq:kernel_vareq}
	\langle \boldsymbol{\phi}(\mathbf{x}(n))|=\sum_{k=1}^p\mathbf{A}_k\langle \boldsymbol{\phi}(\mathbf{x}(n-k))|+\langle \widetilde{\mathbf{w}}(n)|,
\end{equation}
where
\begin{equation}
\label{eq:phi_state_bra}
	\langle \boldsymbol{\phi}(\mathbf{x}(n)) | = [ \langle \phi (x_1(n))|, \cdots, \langle\phi (x_D(n))| ]^{\top},
\end{equation}
which is assumed as due to filtering appropriately modelled innovations $\langle \widetilde{\mathbf{w}}(n)|$. For the present formulation one must also consider the associated `ket'-vector
\begin{equation}
\label{eq:phi_state_ket}
	|\boldsymbol{\phi}(\mathbf{x}(m)) \rangle = [ |\phi (x_1(m))\rangle, \cdots, |\phi (x_D(m))\rangle ]^{\top},
\end{equation}
that when applied to \eqref{eq:kernel_vareq} for $n>m$ after taking expectations $\E[\cdot]$ under the zero mean innovations nature of $\langle \mathbf{w}^\phi(n)|$ leads to
\begin{equation}
	\label{eq:K_yule}
	\mathbf{K}^\phi_\mathbf{x}(l)=\sum_{k=1}^p\mathbf{A}_k \mathbf{K}^\phi_\mathbf{x}(l+k),
\end{equation}
where $l=m-n$ and $\mathbf{K}^\phi_\mathbf{x}(m)$'s elements are given by $\E[\langle \phi(x_i(l-m)) | \phi(x_j(l)) \rangle]$ so that \eqref{eq:K_yule} constitutes a generalization of the Yule-Walker equations. By making $l=m-n=-1, \cdots, -p$ one may reframe \eqref{eq:K_yule} in matrix form as
\begin{equation}
	\label{eq:K_yule_matrix_reduced}
	\bar{\boldsymbol{\kappa}}_p
	=
	\left[
	\begin{array}{c}
		\mathbf{K}^\phi_\mathbf{x}(-1)\\
		\vdots\\
		\mathbf{K}^\phi_\mathbf{x}(-p)
	\end{array}
	\right] 
	= \left[\mathbf{A}_1 \; \cdots \mathbf{A}_p\right]
	\left[
	\begin{array}{cccc}
		\mathbf{K}^\phi_\mathbf{x}(0) 	& \mathbf{K}^\phi_\mathbf{x}(-1) 	& \cdots & \mathbf{K}^\phi_\mathbf{x}(-p+1) \\
		\mathbf{K}^\phi_\mathbf{x}(1) 	& \mathbf{K}^\phi_\mathbf{x}(0)  	& \cdots & \mathbf{K}^\phi_\mathbf{x}(-p+2) \\
		\vdots 						    & \ddots 							& \ddots & \vdots							\\
		\mathbf{K}^\phi_\mathbf{x}(p-1) & \mathbf{K}^\phi_\mathbf{x}(p-2)	& \cdots & \mathbf{K}^\phi_\mathbf{x}(0)
	\end{array}
	\right] 
	= 
	\mathbf{\mathcal{A}}\mathbf{\mathcal{K}}_p(0),
\end{equation}
where $\mathbf{\mathcal{K}}_p(0)$ is block Toeplitz matrix containing $p$ Toeplitz blocks. Equation \eqref{eq:K_yule_matrix_reduced} provides $pD^2$ equations for the same number of unknown parameters in $\mathbf{\mathcal{A}}$.

The high model order counterpart to \eqref{eq:sigma_w} is given by
\begin{equation}
	\label{eq:sigma_w_gen}
	\boldsymbol{\Sigma}_{\langle \widetilde{\mathbf{w}}(n)|}=\mathbf{K}^\phi_\mathbf{x}(0)-\sum_{k=1}^p\sum_{l=1}^p\mathbf{A}_k\mathbf{K}^\phi_\mathbf{x}(k-l)\mathbf{A}_l^{\top}=\mathbf{K}^\phi_\mathbf{x}(0)-\mathbf{\mathcal{A}}\mathbf{\mathcal{K}}_p(0)\mathbf{\mathcal{A}}^\top.
\end{equation}

It is not difficult to see that the more usual Yule-Walker complete equation form becomes
\begin{equation}
	\label{eq:YW_matrix_complete}
	[\mathbf{I} \; -\mathbf{\mathcal{A}}]\; \mathbf{\mathcal{K}}_{p+1}(0)
	=
	\left[
	\begin{array}{c}
		\boldsymbol{\Sigma}_{\langle \widetilde{\mathbf{w}}(n)|}\\
		\mathbf{0}
	\end{array}
	\right].
\end{equation}

There are a variety of ways for solving for the parameters. A simple one is to define $\mathbf{a}=\vect(\mathbf{\mathcal{A}})$ leading to
\begin{equation}
	\label{eq:yule_walker}
	\vect(\bar{\boldsymbol{\kappa}}_p) =(\mathbf{\mathcal{K}}^\top_p(0) \otimes \mathbf{I}) \; \mathbf{a},
\end{equation}
even though one may employ least-squares solution methods to solve \eqref{eq:yule_walker}, or alternatively \eqref{eq:K_yule_matrix_reduced} a Total-least-squares (TLS) approach \cite{Golub2013} is proved a better solution since both members of the equations are affected by estimation inaccuracies that are better dealt with using TLS.

Likewise, \eqref{eq:sigma_w_gen} can be used in conjunction with generalizations of model order criteria of Akaike's AIC type
\begin{equation}
	\label{eq:akaike}
	{_{\mbox{\small{g}}}}{\mbox{AIC}}(k)=\ln (\det(\boldsymbol{\Sigma}_{\langle \widetilde{\mathbf{w}}(n)|}))+\dfrac{c_{n_s}}{n_s}kD^2,
\end{equation}
where $n_s$ stands for the number of available time observations. In generalizing Akaike's criterion to the multivariate case $c_{n_s}=2$, whereas $c_{n_s}=\ln(\ln(n_s))$ for the Hannan-Quinn criterion we use throughout this paper.

So far we described procedures for choosing model order in the $\mathbb{F}$ space. In ordinary time series analysis, in addition to model order identification, one must also perform proper model diagnostics. This entails checking for residual whiteness among other things. This is usually done by checking the residual auto/crosscorrelation functions for their conformity to a white noise hypothesis.  

In the present formulation, because we do not explicitly compute the $\mathbb{F}$ space series, we must resort to means other than computing the latter correlation functions from the residual data as is usual. However, using the same ideas for computing \eqref{eq:sigma_w_gen}, one may  obtain estimates of the innovation cross-correlation in the feature space at various lags as
\begin{equation}
	\label{eq:sigma_w_gen_k}
	\boldsymbol{\Sigma}_{\langle \widetilde{\mathbf{w}}(n)| \widetilde{\mathbf{w}}(m)\rangle}= \boldsymbol{\Sigma}_{\widetilde{\mathbf{w}}}(m-n) = \mathbf{K}^\phi_\mathbf{x}(m-n)-\sum_{k=1}^p\sum_{l=1}^p\mathbf{A}_k\mathbf{K}^\phi_\mathbf{x}(m-n+k-l)\mathbf{A}_l^{\top},
\end{equation}
by replacing $\mathbf{K}^\phi_\mathbf{x}(m-n+k-l)$ by their estimates and using $\mathbf{A}_k$ obtained by solving \eqref{eq:K_yule} for $m-n$ between a minimum $-L$ to a $+L$ maximum lag. The usefulness of \eqref{eq:sigma_w_gen_k} is to provide means to test model accuracy and quality as a function of $\phi$ choice under the best model order provided by the model order criterion.

By defining a suitable normalized estimated lagged \textit{kernel correlation function (KCF)}
\begin{equation}
	\label{eq:kcf}
	\mbox{KCF}_{ij}(\tau) = \dfrac{ K_{ij}(\tau) }{\sqrt{K_{ii}(0) K_{jj}(0)}},
\end{equation}
which, given the inner product nature of kernel definition satisfies the condition
\begin{equation}
	\label{eqkcf_cs}
	| \mbox{KCF}_{ij}(\tau) | \leq 1,
\end{equation}
as easily proved using the Cauchy-Schwarz inequality.

The notion of $\mbox{KFC}(\tau)$ applies not only to the original kernels but also in connection with the residual kernels values given by \eqref{eq:sigma_w_gen_k} where for, for explicitness, we write it as
\begin{equation}
	\label{eq:kcf_}
	\mbox{KCF}^{(r)}_{ij}(\tau) = \dfrac{\Sigma_{ij}(\tau) }{\sqrt{\Sigma_{ii}(0) \Sigma_{jj}(0)}},
\end{equation}
where $\Sigma_{ij}(\tau)$ are the matrix entries in \eqref{eq:sigma_w_gen_k}.

In the numerical illustrations that follow, we have assumed that  $\mbox{KCF}^{(r)}_{ij}(\tau)\sim \mathcal{N}(0,1/n_s)$ asymptotically  under the white residual hypothesis
\begin{equation}
	\label{eq:null_hypo_kcf}
	\mathcal{H}_0: \mbox{KCF}^{(r)}_{ij}(\tau) = 0,
\end{equation}
This choice turned out to be reasonably consistent in practice. In the same line of reasoning, other familiar residual tests over residuals, such as the Portmanteau test \cite{Lutkepohl2005} were also be carried out and managed to reject residual nonwhiteness.

One may say that the present theory follows closely the developments of ordinary second order moment theory with the added advantage that now nonlinear connections can be effectively captured by replacing second order moments by their respective lagged kernel estimates.

\subsection{Estimation and Asymptotic Considerations}
\label{sec:asymp_con}

The essential problem then becomes that of estimating the entries of $\mathbf{K}^\phi_\mathbf{x}(n,m)$, entries. They can be obtained by averaging kernel values computed over the available data
\begin{equation}
	\label{eq:K_matrix_entries}
	K_{ij}(n,m)=\dfrac{1}{n_s}\sum_s \langle \phi(x_i(n-s))|\phi(x_j(m-s))\rangle,
\end{equation}
for nonzero terms in the $s \in [1, n_s]$ range.

Under these conditions, for an appropriately defined kernel function, the feature space becomes linearized and following \cite{Hable2012}, it is fair to assume that the estimated vector stacked  representation of the model coefficient matrices
\begin{equation}
	\label{eq:a_vec}
	\mathbf{a}=\vect([\mathbf{A}_1 \cdots \mathbf{A}_p]) 
\end{equation}
is asymptotically Gaussian, i.e.,
\begin{equation}
\label{eq:a_asymp}
	\sqrt{n_s}(\widehat{\mathbf{a}}-\mathbf{a}) \sim \mathcal{N}(\boldsymbol{0}, \boldsymbol{\Gamma}^{-1} \otimes \boldsymbol{\Sigma}_{\langle \widetilde{\mathbf{w}}(n)|}),
\end{equation}
where $\boldsymbol{\Sigma}_{\langle \widetilde{\mathbf{w}}(n)|}$ is the feature space residual matrix given by \eqref{eq:sigma_w_gen_k} and where
\begin{equation}
	\label{eq:augmented_K}
	\boldsymbol{\Gamma}=\E[\mathbf{y}_L\mathbf{y}^{\top}_R]
\end{equation}
for the `bra'-vector
\begin{equation}
	\mathbf{y}_L^{\top}=[\langle \boldsymbol{\phi}(\mathbf{x}(n))|,\cdots, \langle \boldsymbol{\phi}(\mathbf{x}(n-p+1))| ]^{\top}
\end{equation}
and the `ket'-vector
\begin{equation}
	\mathbf{y}_R^{\top}=[| \boldsymbol{\phi}(\mathbf{x}(n))\rangle, \cdots, | \boldsymbol{\phi}(\mathbf{x}(n-p+1))\rangle ]^{\top},
\end{equation}
that are used to construct the kernel scalar products. It is immediate to note that \eqref{eq:augmented_K} is a Toeplitz matrix composed of suitably displaced $\mathbf{K}_\mathbf{x}^\phi(\cdot)$ blocks.

An immediate consequence of \eqref{eq:a_asymp} is that one may test for model coefficient nullity and thereby provide a kernel Granger Causality test. This is equivalent to testing for $a_{ij}(k)=0$ so that the statistic
\begin{equation}
	\label{eq:kernel_wald}
	{_{\mbox{\small{g}}}}\lambda_W =  \mathbf{\widehat{a}} ^{\top}  \mathbf{C}^{\top} \left[\mathbf{C} \left( \boldsymbol{\Gamma}^{-1}\otimes \boldsymbol{\Sigma}_{\langle \widetilde{\mathbf{w}}(n)|} \right) \mathbf{C}^{\top} \right]^{-1} \mathbf{C} \mathbf{\widehat{a}},
\end{equation}
where $\mathbf{C}$ is a contrast matrix (or structure selection matrix) so that the null hypothesis becomes
\begin{equation}
	\label{eq:null_hypo}
	\mathcal{H}_0: \mathbf{C}\mathbf{a}=0.
\end{equation}

Hence, under \eqref{eq:a_asymp} 
\begin{equation}
	\label{eq:granger_distrib}
	{_{\mbox{\small{g}}}}\lambda_W \stackrel{d}{\to} \chi_{\nu}^2,
\end{equation}
where $\nu = \rank (\mathbf{C})$ corresponds to the number of the explicitly imposed constraints on $a_{ij}(k)$.

%% file: num_ill.tex
The following examples consist of nonlinearly coupled systems that are simulated with help of input zero mean unit variance normal uncorrelated innovations $w_i(n)$. All simulations ($10,000$ realizations each) were preceded by an initial a burn-in period of $10,000$ data points to avoid transient phenomena. Estimation results are examined as a function of $n_s = \{ 32, 64, 128, 256, 512, 1024, 2048 \}$ with $\alpha = 1\%$ significance.

For brevity Example 1 is carried out in full detail whereas approach performance for the other ones is gauged mostly through the computation of observed detection rates except for Examples 4 and 5 which also portray model order choice criterion behaviour.

Simulations results are displayed as function of realization length $n_s$ assuming $\alpha = 1\%$ in testing \eqref{eq:null_hypo}.

\subsection{Example 1}
\label{subsec:example1}

\input{example1}

\subsection{Example 2}
\label{subsec:example2}

\input{example2}

\subsection{Example 3}
\label{subsec:example3}

\input{example3}

\subsection{Example 4}
\label{subsec:example4}

\input{example4}

\subsection{Example 5}
\label{subsec:example5}

\input{example5}

%% file: example1.tex
Consider the simplest possible system whose connectivity cannot be captured by linear methods \cite{Massaroppe2011a} as there is a unidirectional quadratic coupling from $x_2(n)$ to $x_1(n)$
\begin{equation}
	\label{eq:example1}
	\left\{
		\begin{aligned}
			x_1(n) &=            ax_1(n-1) +    cx_2^2(n-1) + w_1(n), \\
			x_2(n) &= \phantom{{}dx_1(n-1) +{}} bx_2  (n-1) + w_2(n),
		\end{aligned}
	\right.
\end{equation}
with $a=0.2$, $b=0.6$ and $c=0.7$.

An interesting aspect of this simple system is the possibility of easily relating its coefficients $a$, $b$ and $c$ to those in \eqref{eq:small_A} that describe its $\mathbb{F}$ space evolution. This may be carried out explicitly after substituting \eqref{eq:example1} into the computed kernels of equation \eqref{eq:kernel_matrix_biv}. After a little algebra, this leads to
\begin{equation}
	\label{eq:little_algebra}
	\left[\left[
	\begin{array}{cc}
		a & 0 \\
		0 & b
	\end{array}\right]-
	\left[
	\begin{array}{cc}
		\alpha_{11} & \alpha_{12} \\
		\alpha_{21} & \alpha_{22} 
	\end{array}\right]\right]
	=\left[
	\begin{array}{cc}
		c & 0 \\
		0 & 0
	\end{array}\right]
	\left[
	\begin{array}{cc}
		\theta_{11} & \theta_{12} \\
		0 & 0
	\end{array}\right],
\end{equation}
where $\theta_{11} $ and $\theta_{12}$ depend on the computed kernel values. From \eqref{eq:little_algebra} it immediately follows for example that $b=\alpha_{22}$ and more importantly that $\alpha_{21}=0$ as expected. Vindication of the observation of these theoretically determined values also gives the means for testing estimation accuracy.

For illustration sake, we write the kernel Yule-Walker equations \eqref{eq:K_yule} with their respective solutions ($n_s=512$) for one given (typical) realization
\begin{equation}
	\label{eq:est_sec_power}
	\left[
	\begin{array}{cc}
		210.7583 & 23.5416 \\
		 23.5416 &  8.6450
	\end{array}\right] {\mathbf{A}^{(2)}} =
	\left[
	\begin{array}{cc}
		125.7501 & 37.7803 \\
		 17.7389 &  5.3788
	\end{array}\right] \rightarrow {\mathbf{A}^{(2)}} =
	\left[
	\begin{array}{cc}
		0.1559 & 3.9456 \\
		0.0211 & 0.5648
	\end{array}\right],
\end{equation}
for the quadratic kernels ($\kappa(x,y) = (xy) ^2$) and
\begin{equation}
	\label{eq:est_fourth_power}
	10^5 \times \left[
	\begin{array}{cc}
		8.0302 & 0.0868 \\
		0.0868 & 0.0052
	\end{array}\right] {\mathbf{A}^{(4)}} =
	10^5 \times \left[
	\begin{array}{cc}
		4.1597 & 0.1755 \\
		0.0594 & 0.0025
	\end{array}\right] \rightarrow {\mathbf{A}^{(4)}} =
	\left[
	\begin{array}{cc}
		0.1843 & 30.8922 \\
		0.0027 &  0.4386
	\end{array}\right],
\end{equation}
for the quartic kernels ($\kappa(x,y) = (xy) ^4$). Superscripts point to kernel order. One may readily notice approximate compliance to the expected $\alpha_{ij}$ coefficients.

\begin{figure}[htbp!]
\label{fig:initial_kerneL_corre}
	\centering
	\subfloat[$\kappa(x,y) = (xy) ^2$\label{fig:kernel_seq_kxcorr2}]{%
		\includegraphics[scale=.45]{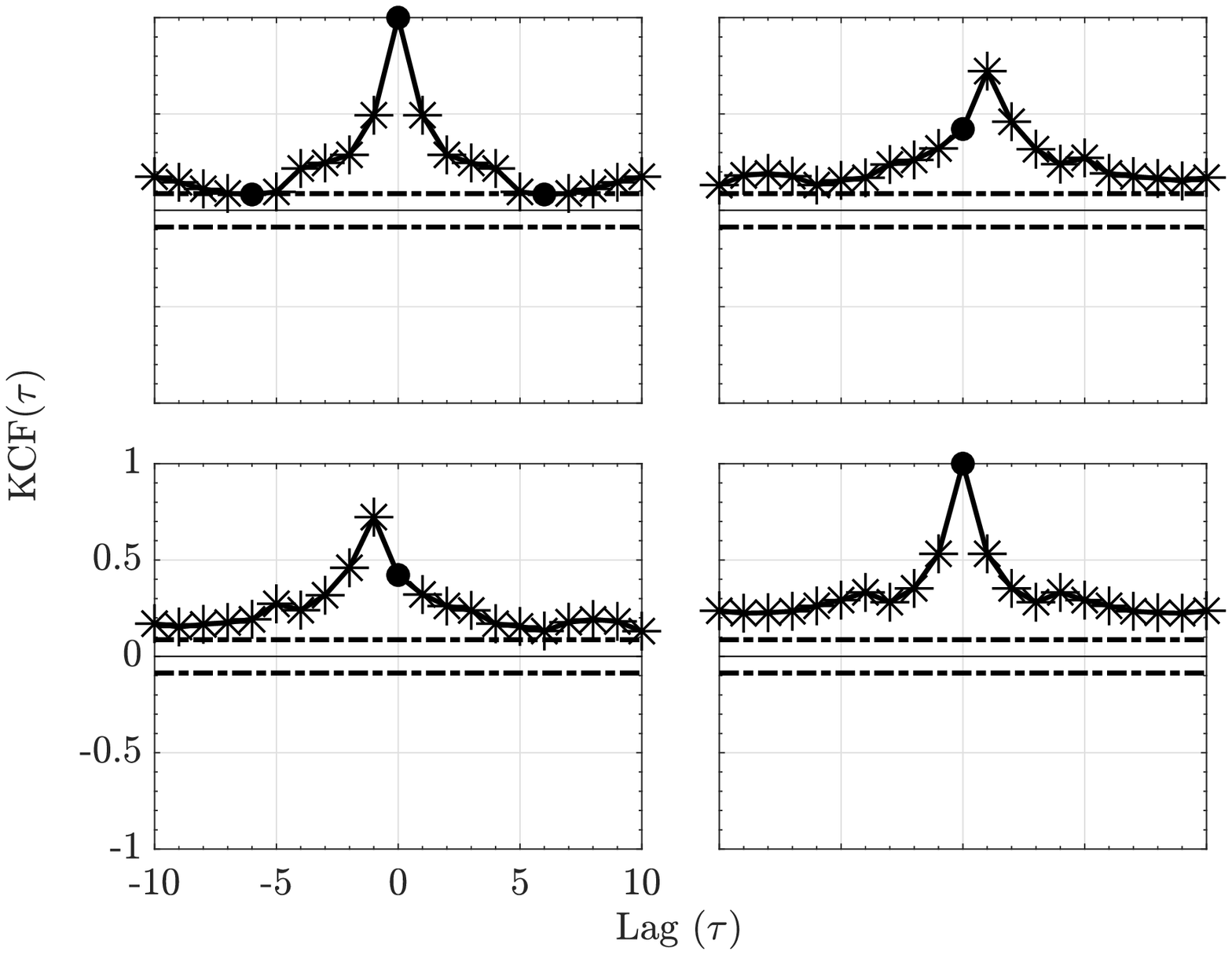}
	}
	\hspace*{10pt} 
	\subfloat[$\kappa(x,y) = (xy) ^4$.\label{fig:kernel_seq_kxcorr4}]{%
		\includegraphics[scale=.45]{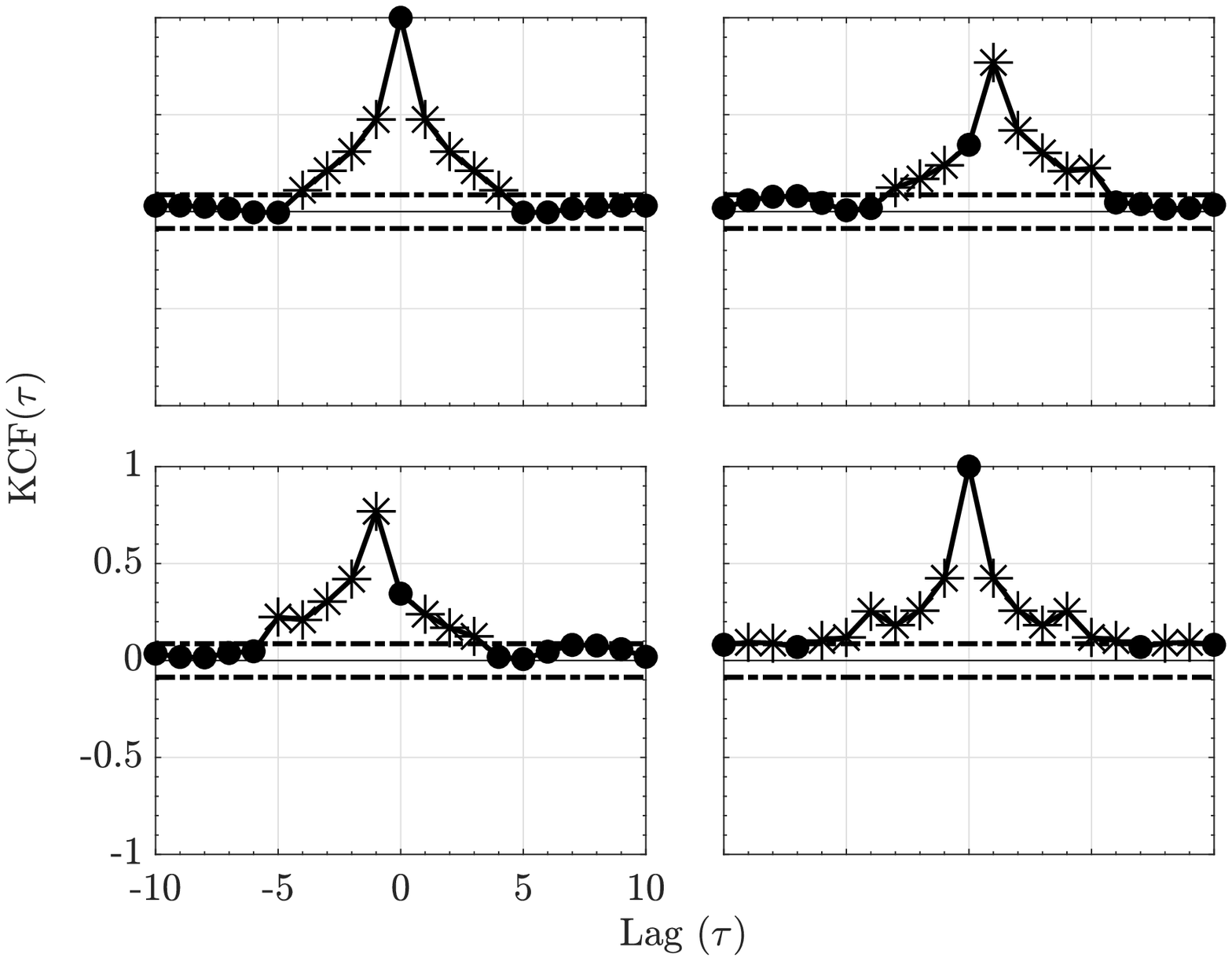}
	}
	\caption{The sequence kernel correlation functions (KCF$(\tau)$) for both kernels depicted in Figures (\ref{fig:kernel_seq_kxcorr2}) and (\ref{fig:kernel_seq_kxcorr4}) for Example 1.}
	\label{fig:kernel_seq_kxcorr}
\end{figure}

Further appreciation of this example may be obtained via a plot of the normalized estimated lagged kernel sequences \eqref{eq:kcf} shown in Figure \ref{fig:kernel_seq_kxcorr}. 

\begin{figure}[htbp!]
\label{fig:residual_kernel_sequences}
	\centering
	\subfloat[$\kappa(x,y) = (xy) ^2$\label{fig:residue_xcorr_2}]{%
		\includegraphics[scale=.45]{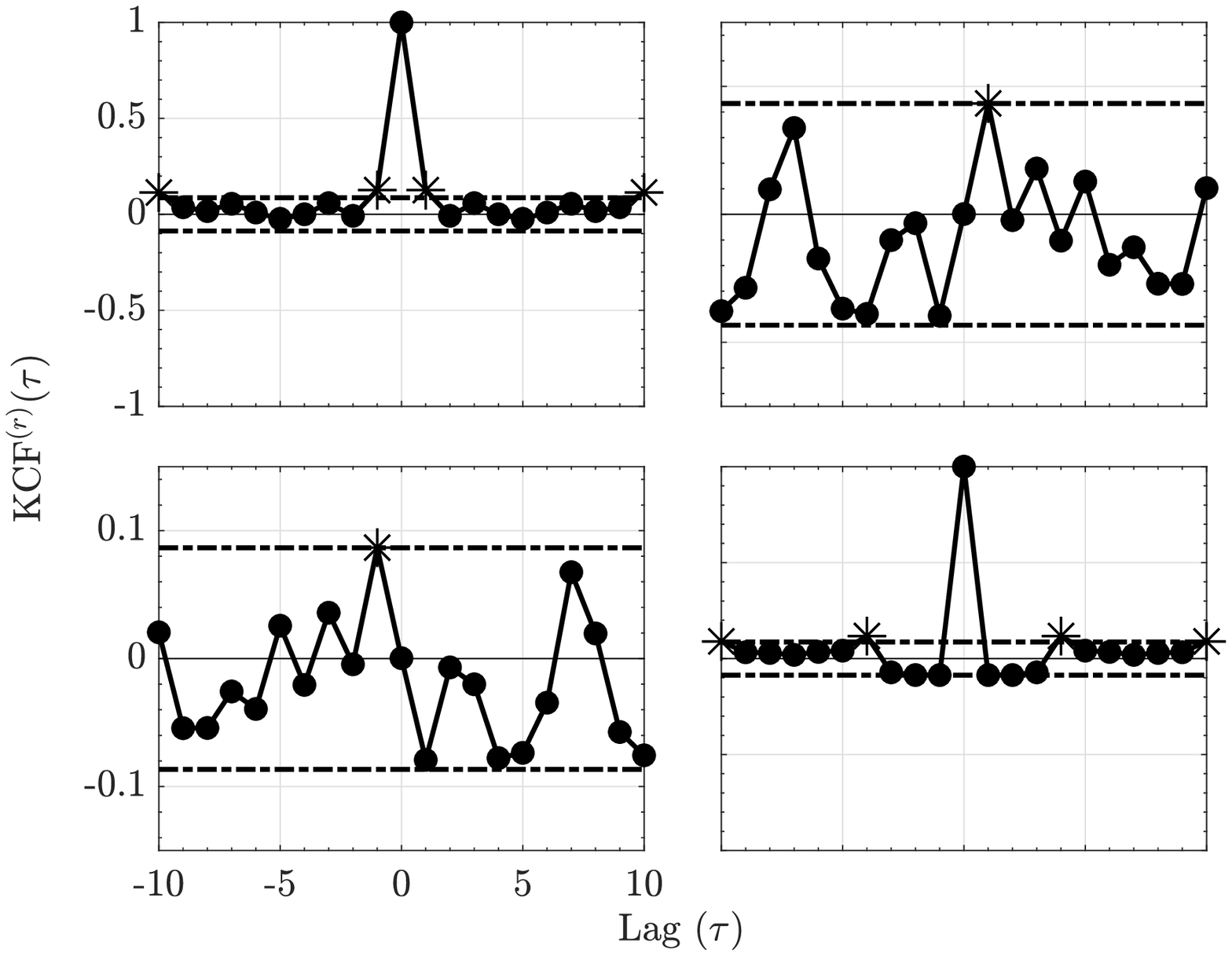}
	}
	\hspace*{10pt} 
	\subfloat[$\kappa(x,y) = (xy) ^4$.\label{fig:residue_xcorr_4}]{%
		\includegraphics[scale=.45]{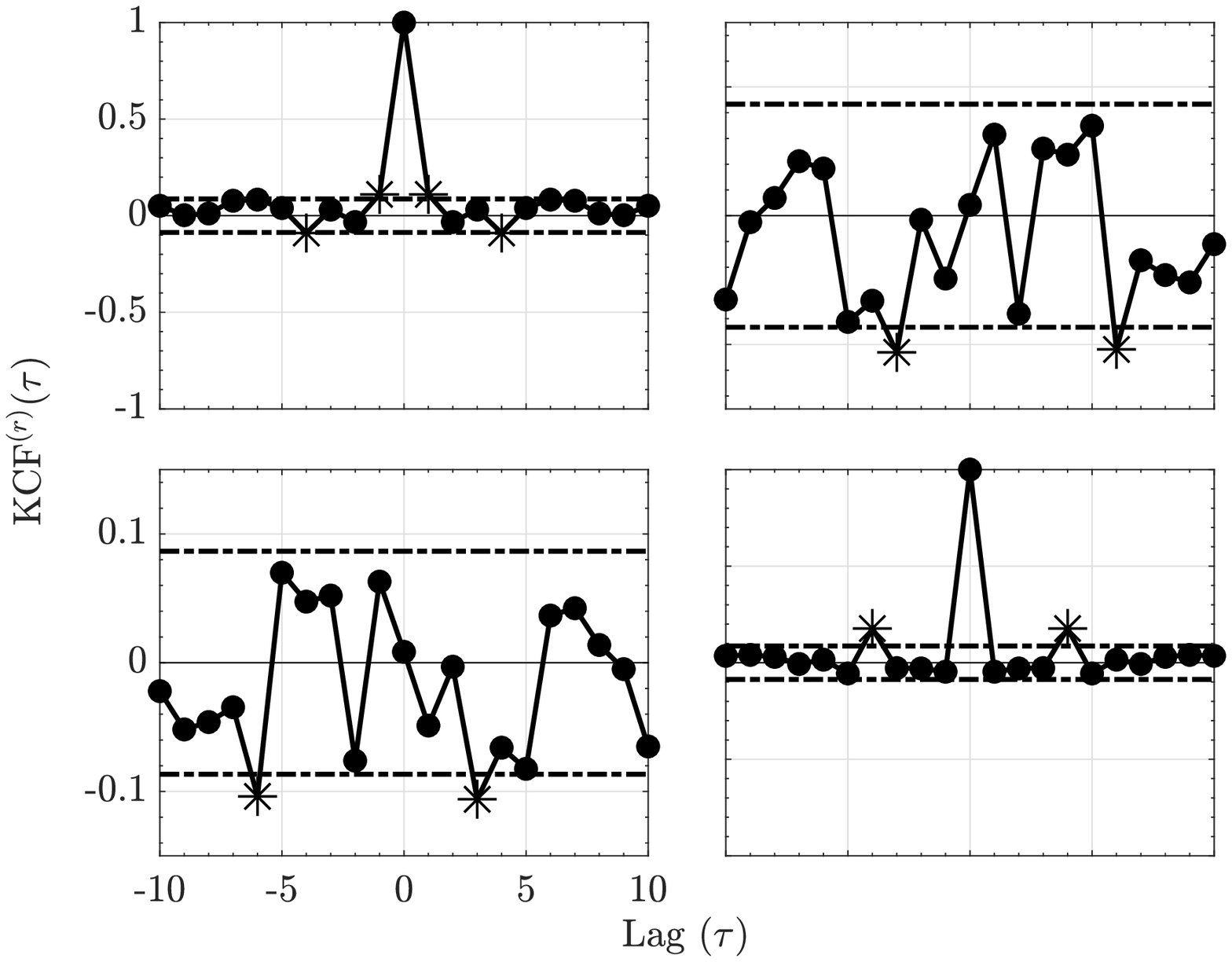}
	}
	\caption{The residue kernel correlation functions (KCF$^{(r)}(\tau)$) for both kernels depicted in Figures (\ref{fig:residue_xcorr_2}) and (\ref{fig:residue_xcorr_4}) for Example 1.}
	\label{fig:residue_plot}
\end{figure}

The residual normalized kernel sequences \eqref{eq:kcf_} computed using \eqref{eq:sigma_w_gen_k} are depicted in Figure \ref{fig:residue_plot} for each kernel and show effective decrease below the null hypothesis decision threshold line vindicating adequate modelling. 

Moreover, for this realization, one may show that the Hannan-Quinn Information Criterion \eqref{eq:akaike} points to the correct order of $p=1$. Also, Portmanteau tests do not reject whiteness in the $\mathbb{F}$ space for either kernel further confirming successful modelling in both cases.

\begin{figure}[htbp!]
	\centering
	\subfloat[$\kappa(x,y) = (x,y) ^2$\label{fig:normplot_a21_m_211}]{%
		\includegraphics[scale=.45]{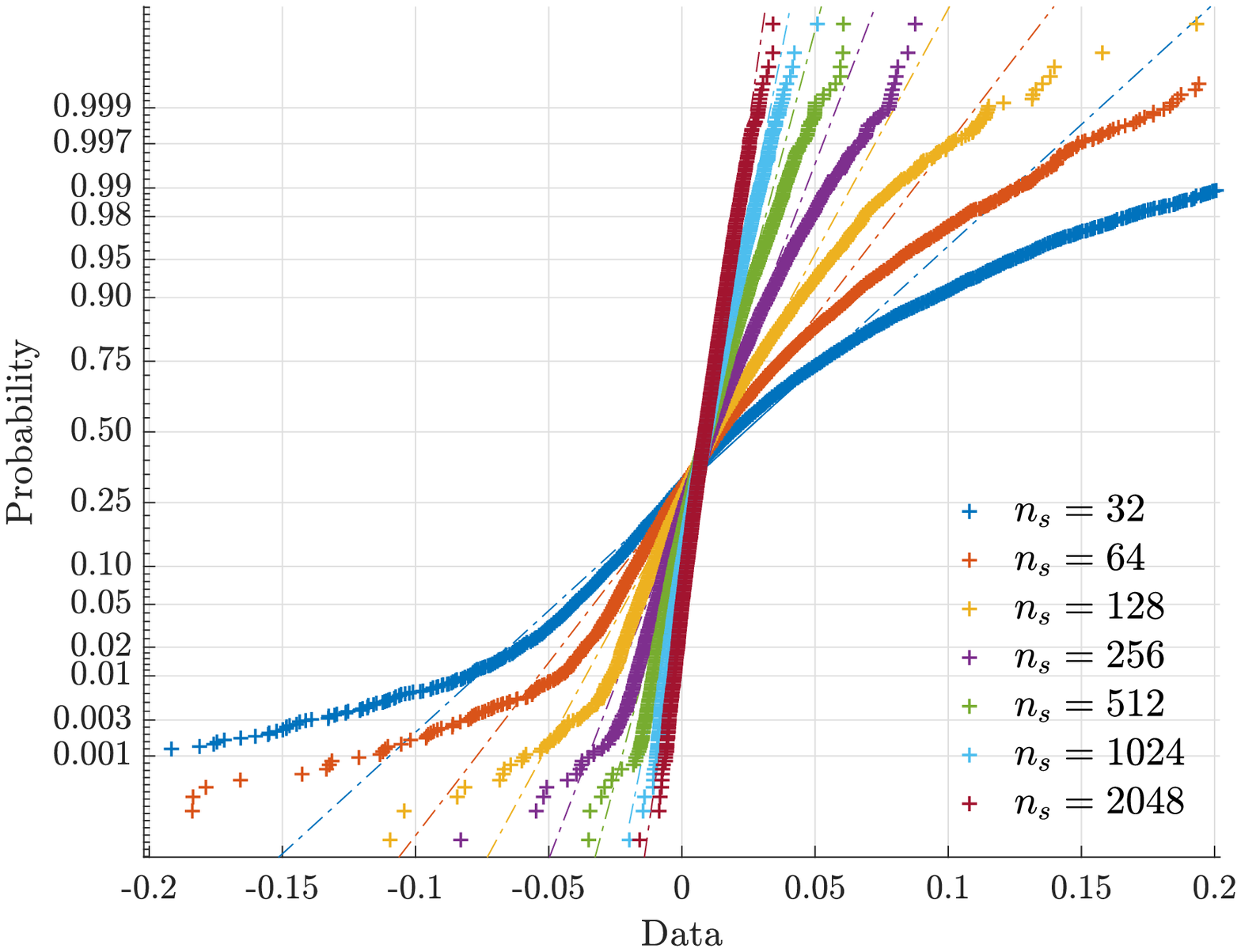}
	}
	\hspace*{10pt} 
	\subfloat[$\kappa(x,y) = (x,y) ^4$.\label{fig:normplot_a21_m_411}]{%
		\includegraphics[scale=.45]{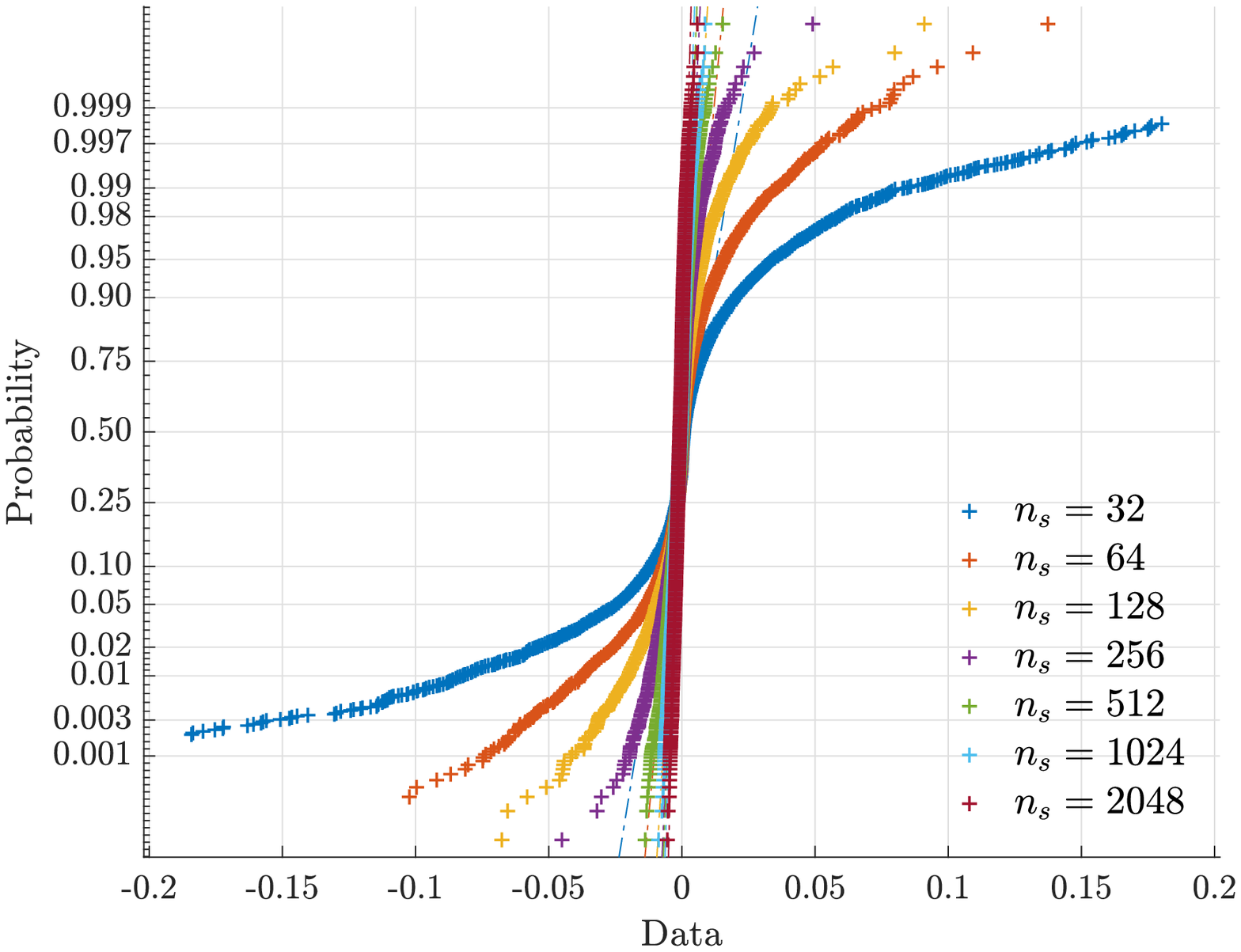}
	}
	\caption{Ensemble normal probability plots for $\widehat{a}_{21}$, respectively for (\ref{fig:normplot_a21_m_211}) quadratic and (\ref{fig:normplot_a21_m_411}) quartic kernels illustrate and confirm asymptotic normality.}
	\label{fig:normplot}
\end{figure}

\begin{figure}[htbp!]
	\centering
	\includegraphics[scale=.45]{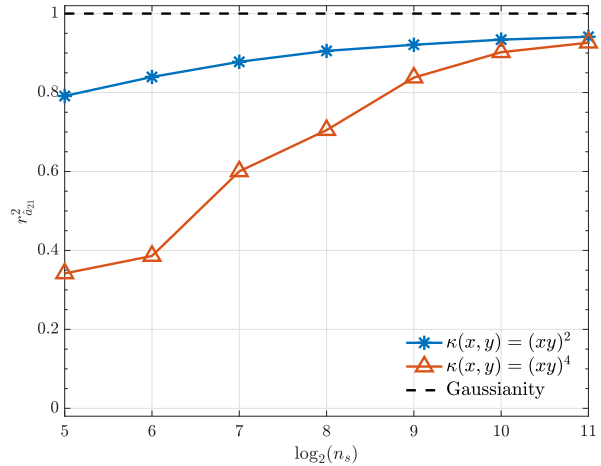}
	\caption{\emph{Filliben} probability plot squared-correlation coefficient evaluated for both kernels.}
	\label{fig:gof_example1}
\end{figure}

To illustrate and confirm the Gaussian asymptotic behaviour discussed in Section \ref{sec:asymp_con}, normal probability plots for $\widehat{a}_{21}$ are presented in Figure \ref{fig:normplot}. Further objective quantification of the convergence speed towards normality is provided by the evolution towards $1$ of the \emph{Filliben} squared-correlation coefficient \cite{Filliben1975, Vogel1986, Vogel1987_Corrections} as a function of $n_s$ (Figure \ref{fig:gof_example1}).

\begin{figure}[htbp!]
	\centering
	\includegraphics[scale=.45]{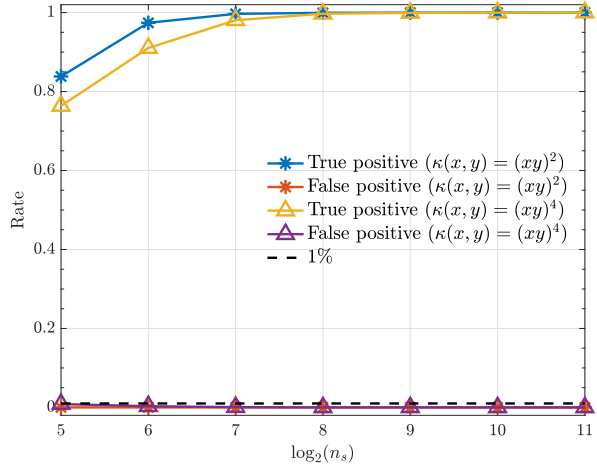}
	\caption{True positive and false positive rates from the kernelized Granger causality test for various samples sizes ($n_s$) for $\alpha=1\%$.}
	\label{fig:tp_fp_example1}
\end{figure}

Convergence to normality justifies using \eqref{eq:kernel_wald} to test for null connectivity hypotheses. Test perfomance is depicted in Figure \ref{fig:tp_fp_example1}.

%% file: example2.tex
Consider a highly resonant ($R=0.99$) linear oscillator $x_1(n)$  unidirectionally coupled to a low pass system $x_2(n)$ through a delayed squared term
\begin{equation}
	\label{eq:example2}
	\left\{
		\begin{aligned}
			x_1(n) &= 2 R \cos(2 \pi f) x_1(n - 1) - R^2 x_1(n - 2) + w_1(n), \\
			x_2(n) &= -0.9 x_2(n - 1) + c x_1^{2}(n - 1) + w_2(n),
		\end{aligned}
	\right.
\end{equation}
where $c=0.1$ \cite{Massaroppe2011a}.

\begin{figure}[htbp!]
	\centering
	\vspace{.1in}
	\includegraphics[scale=.45]{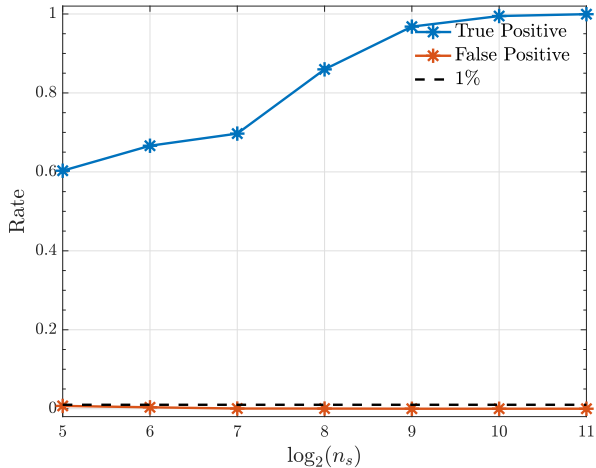}
	\caption{True positive and false positive rates from the kernelized Granger causality test under a quadratic kernel as a function $n_s$ in Example 2.}
	\label{fig:tp_fp_example2}
\end{figure}

This system was already investigated elsewhere \cite{Massaroppe2011a, Massaroppe2011b, Massaroppe2015a} under a different estimation algorithm and with fewer Monte Carlo replications. The null hypothesis connectivity results are presented in Figure \ref{fig:tp_fp_example2} showing adequate asymptotic decision success. A quadratic kernel was used in all cases.

%% file: example3.tex
The present example comes from a model in \cite{Gourevitch2006},
\begin{equation}
	\label{eq:example3}
	\left\{
		\begin{aligned}
			x_1(n) &= 3.4 x_1(n-1)[1 - x_1^{2}(n-1)]\e^{-x_1^{2}(n-1)} \phantom{{}+ c_3 x_1^{2}(n-1){}} + w_1(n), \\
			x_2(n) &= 3.4 x_2(n-1)[1 - x_2^{2}(n-1)]\e^{-x_2^{2}(n-1)}            + c_1 x_1^{2}(n-1)    + w_2(n), \\
			x_3(n) &= 3.4 x_3(n-1)[1 - x_3^{2}(n-1)]\e^{-x_3^{2}(n-1)}            + c_2 x_2^{4}(n-1)    + w_3(n).
		\end{aligned}
	\right.
\end{equation}
This choice was dictated by the nonlinear wideband character of its signals. The values $c_1 = 0.7$ and $c_2 = 0.9$ were adopted.

\begin{figure}[htbp!]
	\centering
	\includegraphics[scale=.45]{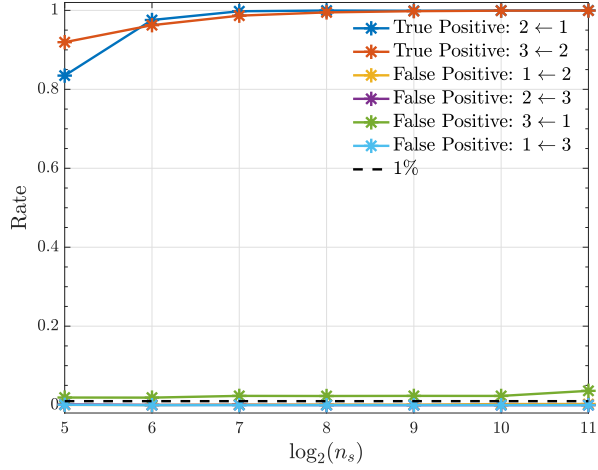}
	\caption{True positive and false positive rates from the kernelized Granger causality test using a quadratic kernel as a function of $n_s$.}
	\label{fig:tp_fp_example3}
\end{figure}

Figure \ref{fig:tp_fp_example3} shows that connection detectability improves as signal duration $n_s$ increases except for the nonexisting $x_3(n) \leftarrow x_1(n)$ connection whose performance stays more or less constant with a false positive rate slightly above $\alpha=1\%$. All computations used quadratic kernels.

%% file: example4.tex
For this numerical illustration, consider the model presented in \cite{Chen2004}
\begin{equation}
	\label{eq:example4}
	\left\{
		\begin{aligned}
			x_1(n) &= 3.4 x_1(n-1)[1 - x_1^{2}(n-1)]\e^{-x_1^{2}(n-1)} + 0.8 x_1(n-2) \phantom{{}+ c x_2^{2}(n-2){}} + w_1(n), \\
			x_2(n) &= 3.4 x_2(n-1)[1 - x_2^{2}(n-1)]\e^{-x_2^{2}(n-1)} + 0.5 x_2(n-2)            + c x_1^{2}(n-2)    + w_2(n).
		\end{aligned}
	\right.
\end{equation}
System \eqref{eq:example4} produces nonlinear wideband signals with a quadratic ($1 \leftarrow 2$) coupling factor whose intensity is given by $c$ taken here as $0.5$.

\begin{figure}[htbp!]
	\centering
	\includegraphics[scale=.45]{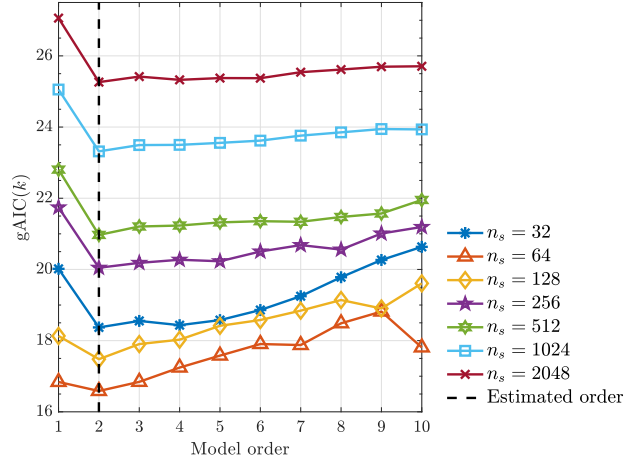}
	\caption{Generalized Hannan-Quinn criterion ($ _{\mbox{\small{g}}}{\mbox{AIC}}(k)$)  with $c_{n_s}=\ln(\ln (n_s))$ as a function of model order for various observed record lengths $n_s$ using a typical realization from \eqref{eq:example4}.}
	\label{fig:info_crit_analysis_example4}
\end{figure}

It is worth noting that, using the generalized Hannan-Quinn criterion, the order of kernelized autoregressive vector models identified for a typical realization was correctly identified and equals 2 as expected (See Figure \ref{fig:info_crit_analysis_example4}).
\begin{figure}[htbp!]
	\centering
	\includegraphics[scale=.45]{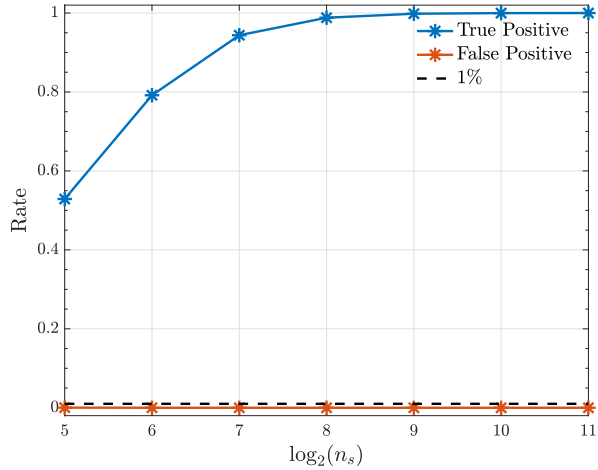}
	\caption{Observed true positive and false positive rates from the kernelized Granger causality test under a quadratic kernel as function of record length $n_s$ in Example 4.}
	\label{fig:tp_fp_example4}
\end{figure}

%% file: example5.tex
As a last numerical illustration, consider data generated by
\begin{equation}
	\label{eq:example5}
	\left\{
		\begin{aligned}
			x_1(n) &= 3.4 x_1(n-3)[1 - x_1^{2}(n-3)]\e^{-x_1^{2}(n-3)}            + 0.4 x_1(n-4)    \phantom{{}+ c_3 x_1^{2}(n-1){}} + w_1(n), \\
			x_2(n) &= 3.4 x_2(n-1)[1 - x_2^{2}(n-1)]\e^{-x_2^{2}(n-1)} \phantom{{}+ 0.0 x_1(n-1){}}            + c_1 x_1^{2}(n-2)    + w_2(n), \\
			x_3(n) &= 3.4 x_3(n-2)[1 - x_3^{2}(n-2)]\e^{-x_3^{2}(n-2)} \phantom{{}+ 0.0 x_1(n-1){}}            + c_2 x_2^{2}(n-3)    + w_3(n),
		\end{aligned}
	\right.
\end{equation}
with $c_1=0.9$ and $c_2=0.4$.

\begin{figure}[htbp!]
	\centering
	\includegraphics[scale=.45]{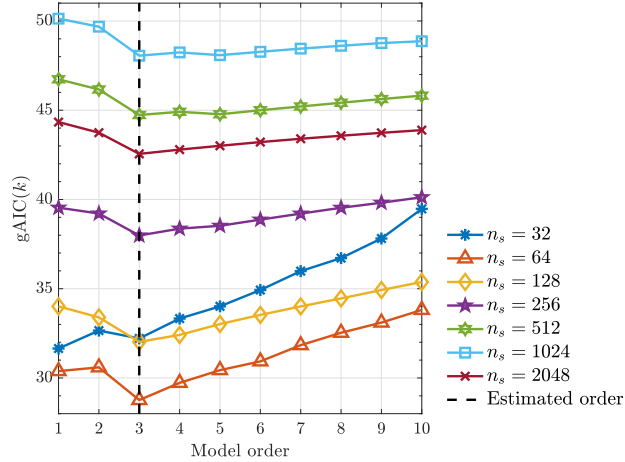}
	\caption{Generalized Hannan-Quinn criterion ($_{\mbox{\small{g}}}{\mbox{AIC}}(k)$) as a function of model order for the various data lengths $n_s$ from a typical realization from \eqref{eq:example5}.}
	\label{fig:info_crit_analysis_example5}
\end{figure}

\begin{figure}[htbp!]
	\centering
	\includegraphics[scale=.45]{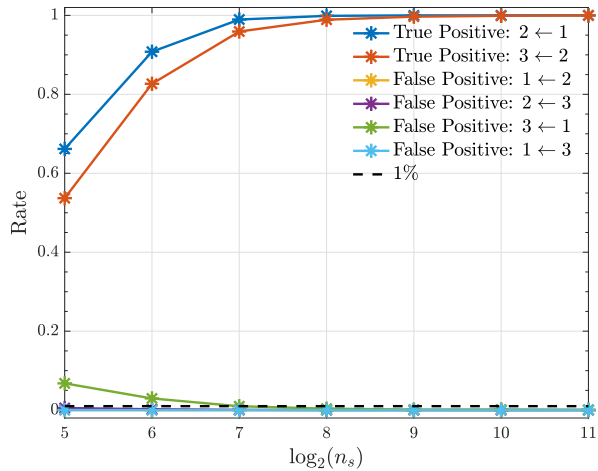}
	\caption{Observed true positive and false positive rates from the kernelized Granger causality test under a quadratic kernel for various record lengths $n_s$ in Example 5.}
	\label{fig:tp_fp_example5}
\end{figure}

Under the quadratic kernel and employing kernelized Hannan-Quinn information criterion \eqref{eq:akaike} (see Figure \ref{fig:info_crit_analysis_example5}) one can see that the estimated model order is $p=3$ as expected judging from the $x_2^2(n-3)$ term in \eqref{eq:example5}. Also kernelized Granger causality detectability improves with record length $n_s$ increase (Figure \ref{fig:tp_fp_example5}).

%% file: disc.tex
After a brief theoretical presentation (Section \ref{sec:prob_form}) we have shown that canonical model fitting procedures that involve (a) model specification with order determination and (b) explicit model diagnostic testing can be successfully carried out in the feature space $\mathbb{F}$ to detect connectivity via reproducing kernels. The key results behind (b) is \eqref{eq:sigma_w_gen_k} and the realization that kernel quantities may be normalized much as correlation coefficients.

What importantly sets the present approach apart is the lack of need for returning to the original input space $\mathbb{X}$ to gauge model quality thereby circumventing the so-called \textit{reconstruction/pre-image problem} \cite{Honeine2011} that may introduce unnecessary uncertainties. We showed that because model adequacy testing can be performed \textit{directly} in the feature space $\mathbb{F}$ directional Granger type connectivity can be detected for a variety of multivariate nonlinear coupling scenarios thereby totally dispensing with the need for detailed `a priori' model knowledge. Thus successful connectivity detection is achievable at the expense of relatively short time series. A systematic comparison with other approaches is planned for future work.

One of the basic tenants of the present work is that model coefficients in the feature space are asymptotically normal, something whose consistency was successfully illustrated though the need for a more formal proof remains especially in connection to explicit kernel estimates under the total-least-squares solution to \eqref{eq:K_yule_matrix_reduced}. Our choice of TLS was dictated by its apparent superiority when compared to the `kernel trick' \cite{Kumar2007} whose multivariate version we employed in \cite{Massaroppe2015a, Massaroppe2015c, Massaroppe2016a}.

Even though order estimation and model testing were successful upon borrowing from the usual linear modelling practices, a further systematic examination is still needed and is underway.

One may rightfully argue that the kernels we chose for illustrating the present work are equivalent to modelling the original time series after the application of a suitable $\phi(\cdot)$ transformation to the data and that they look for the causality evidence present in higher order momenta. This, in fact, explains why quadratic kernels converge much faster than quartic ones in Example 1. The merit of framing the time series transformation discussion for connectivity detection in terms of kernels produces a simple workflow and paves the way to developing future data-driven criteria towards optimum data transformation choice for a given problem. Other kernel choices are being investigated.

One of the advantages of the present development is that the procedure allows determining how far in the past to look via the model order criteria we employed \eqref{eq:akaike}.

Finally, the present systematic empirical investigation sets the proposal of using feature space-frequency domain descriptions of connectivity like \textit{kernel partial directed coherence} \cite{Massaroppe2015a, Massaroppe2015c}, and \textit{kernel directed transfer function} \cite{Massaroppe2016a} on sound footing especially in respect to their asymptotic connectivity behaviour.